\documentclass[fleqn,usenatbib]{mnras}

\usepackage[T1]{fontenc}

\DeclareRobustCommand{\VAN}[3]{#2}
\let\VANthebibliography\thebibliography
\def\thebibliography{\DeclareRobustCommand{\VAN}[3]{##3}\VANthebibliography}

\usepackage{graphicx}	
\usepackage{amsmath}	
\usepackage{amssymb}	
\usepackage{scalefnt}
\usepackage{xspace}

\usepackage{newtxtext,newtxmath}

\def\lsim{\;\raise0.3ex\hbox{$<$\kern-0.75em\raise-1.1ex\hbox{$\sim$}}\;}
\def\gsim{\;\raise0.3ex\hbox{$>$\kern-0.75em\raise-1.1ex\hbox{$\sim$}}\;}

\def \kms {\rm ~km~s$^{-1}$}

\def\ergs{\rm ~erg~s$^{-1}$}

\def\efl{\hbox{erg cm$^{-2}$ s$^{-1}$} }

\newcommand{\cmc}{\,cm$^{-3}$\,}

\newcommand{\ergcc}{\,erg\,cm$^{-3}$\,}
\newcommand{\Msun}{\,M$_{\odot}$\,}

\def\chandra{{\it Chandra}\xspace}
\def\art{{ART-XC}\xspace}
\def\sart{{\it SRG}/ART-XC\xspace}
\def\srg{{\it SRG}\xspace}

\title[Westerlund 2 with \sart and \chandra]{X-ray emission from Westerlund 2 detected by {\bfseries \itshape SRG}/ART-XC  and {\bfseries \itshape Chandra}: search for radiation of TeV leptons} 

\author[A. M. Bykov et al.]{A. M. Bykov,$^{1}$\thanks{E-mail: byk@astro.ioffe.ru } Yu. A. Uvarov,$^{1}$  M. E. Kalyashova,$^{1}$ D. V. Badmaev,$^{1}$ I. Yu. Lapshov,$^{2}$ A. A. Lutovinov,$^{2}$ \newauthor{I. A. Mereminskiy,$^{2}$ A. N. Semena$^{2}$} \\
$^{1}$Ioffe Institute, 26 Politekhnicheskaya St, 194021 Saint-Petersburg, Russia \\
$^{2}$Space Research Institute of RAS, Profsoyuznaya str. 84/32, 117997 Moscow, Russia
}

\date{Accepted XXX. Received YYY; in original form ZZZ}

\pubyear{2023}

\begin{document}
\label{firstpage}
\pagerange{\pageref{firstpage}--\pageref{lastpage}}
\maketitle

\begin{abstract}
We present the results of current observations of the young compact cluster of massive stars Westerlund 2 with the Mikhail Pavlinsky \art\ telescope aboard the Spectrum-Roentgen-Gamma (\srg) observatory which we analysed together with the archival \chandra\ data. In general, Westerlund 2 was detected over the whole electromagnetic spectrum including high-energy gamma rays, which revealed a cosmic ray acceleration in this object to the energies up to tens of TeV. The detection of Westerlund 2  with \art\  allowed us to perform a joint spectral analysis together with the high resolution \chandra\ observations of the diffuse emission from a few selected regions in the vicinity of the Westerlund 2 core in the 0.4 - 20 keV range. To fit the Westerlund 2 X-ray spectrum above a few keV one needs either a non-thermal power-law emission component, or a hot plasma with temperatures $\sim$ 5 keV. Our magnetohydrodynamic modeling of the plasma flows in  Westerlund 2 shows substantially lower electron temperatures in the system and thus the presence of the non-thermal component is certainly preferable. A kinetic model of the particle acceleration demonstrated that the non-thermal component may originate from the synchrotron radiation of multi-TeV electrons and positrons produced in Westerlund 2 in accordance with the TeV photons detection from the source.            
\end{abstract}




\section{Introduction}
\label{sec:intro}
Young clusters of massive stars are produced in the hierarchical star formation process.  Their dynamical evolution and impact on the galactic environment are subjects of great interest \citep[see e.g.][]{Zinn07,Kru19,Ada20,Kra20}. The Hubble imaging Probe of Extreme Environments and Clusters survey revealed the efficient cluster formation in merging galaxies with high star-forming activity typical for the cosmic noon \citep[see e.g.][]{ada20MN}, while the young clusters are less frequent in the Milky Way \citep[e.g.][]{PZ10}. Objects of this type are rare in the Galaxy but provide excellent laboratories to study the complex processes under extreme gas physical conditions in clusters \citep{Clark18}. 

A few of the most massive young galactic clusters, such as Westerlund 1, Westerlund 2 (Wd2 hereafter), Arches and Quintuplet were observed over the whole electromagnetic wave band in the past decade. A notable signature of the complex plasma dynamics in compact cluster cores containing very luminous stars with colliding powerful stellar winds is the detection of very high energy (VHE) gamma rays from the vicinity of Westerlund 1 and 2 \citep{Wd2_Ahar07,Ohm_clusters10,Wd2_Ahar11,Wd1_HESS12,Wd1_HESS22,Wd2_Yang18,Wd2_Mestre21}. The detection of TeV regime photons indicates the presence of particle acceleration process likely powered by shocks produced by the stellar winds and possibly by supernovae \citep{Byk14,Byk20,Morlino21,Bhadra22,Gabici23,Vieu23}. Moreover, compact clusters of massive stars can be considered as important sources of the bulk of the  population of high-energy galactic cosmic rays \citep[see, e.g.,][]{Aha19} among other possibilities \citep[see, e.g.,][]{Amato21}. High resolution radio and X-ray observations are used to study the plasma processes in compact stellar clusters and to understand the origin of non-thermal components. In particular, \citet{Kavanagh20} presented a review of thermal and non-thermal X-ray emission of massive stellar clusters and superbubbles in connection with their evolution status. 

RCW 49 is a giant HII and star formation region, associated with the massive cluster Westerlund 2. Early X-ray observations of Wd2 and RCW 49 with the {\sl Einstein} observatory \citep{Wd2_Einstein_Goldwurm87} and  {\sl ROSAT} telescope \citep{Wd2_ROSAT94}  revealed the peak in its core region and resolved at least three point sources.
Then observations of the RCW 49 region and Wd2 cluster with \chandra\  \citep{Townsley2005,Wd2_Townsley19,Naze08} allowed investigating in great details the hot plasma in early type massive stars. Apart from a dozen of O3 to O6.5 stars, a very interesting eclipsing binary system WR20a in the core of Wd2 was studied with \chandra\ \citep{Naze08}. Two very massive WR stars in WR20a demonstrate the colliding wind system where a hot plasma reaches the temperatures of $\sim 2~\mathrm{keV}$. The massive star WR20b is slightly less luminous and located outside the Wd2 core. It was found that, while most of the massive stars in general follow the relation between the X-ray and bolometric luminosity $L_X \sim 10^{-7} L_{bol}$,  there are massive stars where the ratio is an order of magnitude higher. \chandra\ data analysis of Wd2 by \citet{Townsley2005} established a presence of the diffuse X-ray emission spreading over several-parsec scales with a broad range of plasma temperatures (T $\sim 0.1 - 7$ keV) and possibly even non-thermal emission with a power-law photon indices of $\sim$ 2.

Analysing {\sl Suzaku} observations of Wd2, \citet{Wd2_Suzaku09}, same as \citet{Townsley2005}, came to the conclusion that the X-ray emission can be described with either three thermal components, or two thermal and one non-thermal components. The authors suggested that the active particle acceleration has stopped in Wd2 and that the observed gamma-ray emission is possibly produced by the previously accelerated  high-energy protons. They also noted that the derived  metal abundances of the diffuse X-ray gas may be provided by a core-collapse supernova in the past.

A more massive and older \citep[see, e.g.,][for a recent discussion]{Bea21} cluster Westerlund 1   was studied in X-rays with {\sl Chandra} by \citet{Mun06} who revealed that its diffuse emission extends far beyond the cluster core radius.   
After the subtraction of all the point-like sources  \citet{Mun06} identified a diffuse 2--8 keV emission 
with the luminosity of $(3 \pm 1)\times 10^{34}$ \ergs in the $\sim5^\prime$ cluster halo. 
The diffuse emission of Westerlund 1 is dominated by a hard component which may be attributed either to the emission of a thin thermal plasma of $\sim$ 3 keV temperature with the reduced iron abundance (about 0.3 of solar) or to the non-thermal emission 
with a power-law photon index of $\sim$ 2 (outside the central circle of $1^\prime$ radius).
{\sl XMM-Newton} observations of Westerlund 1 \citep{KavanaghWd1} demonstrated  
a presence of a prominent He-like Fe 6.7 keV line in the X-ray spectrum of the central circle of $2^\prime$ radius which indicated the thermal origin of the hard emission of the cluster core. 
\citet{Mun06} noted that if the non-thermal component of the halo is real, it can be due to an inverse Compton (IC) scattering of the optical radiation by mildly relativistic electrons 
\citep[see also][]{Wd2_Bedanarek97}. The IC origin of the observed very high energy gamma-ray emission in Westerlund 1 within a model of CR acceleration by the cluster wind shock was advocated recently by \citet{Wd1_Harer23}. Indeed, in this model the plasma surrounding of the extended cluster wind shock is dilute and the energetic efficiency of the IC radiation is higher than that in the hadronic processes.

Other compact stellar clusters were detected in X-rays as well. A \chandra\ detection of an early type stars emission in the Tarantula Nebula in LMC \citep{Tarantula_Chandra22} provided a way to establish the average plasma temperature of luminous O-stars in the nebula to be $\sim 1~\mathrm{keV}$, while in the Carina region of the Milky Way it is $\sim 0.5~\mathrm{keV}$. The massive stellar cluster RMC 136 in LMC was recently observed in X-rays with \srg/eROSITA by \citet{Sasaki22} and some hints of a non-thermal component with a hard power-law index were found. Also, {\it NuSTAR} and {\it XMM-Newton} observations of the non-thermal continuum emission and the Fe 6.4 keV line flux variability in the galactic center cluster Arches were discussed by \citet{Arches_Nustar19}. 

Thus, the likely presence of the non-thermal emission was found in a few clusters including the high-energy gamma-ray sources Westerlund 1 and 2. The origin of the non-thermal X-ray component is a subject of interest. In this paper we report the detection of the X-ray emission from the massive star cluster Wd2 with the grazing-incidence Mikhail Pavlinsky \art\ X-ray  telescope \citep{ARTXC21} operating in the 4-30 keV energy range aboard the Spectrum-Roentgen-Gamma (\srg) observatory \citep{SRG21}. These data are analysed together with the archival \chandra\ data. The unique spatial resolution of \chandra\ allows us to distinguish different emission components in the very compact cluster containing many X-ray sources.  

The paper is organized as follows. In \S \ref{sec:multi} we review the multiwavelength observations of RCW 49 and Wd2. In \S \ref{sec:obs} we present the results of new \art\ observations of the Wd2 field. Joint analysis of X-ray \art\ and archival \chandra\ observations of the vicinity of Wd2 and the results of the 3D MHD modeling of plasma flows in the Wd2 cluster are presented in \S \ref{sec:an}. In \S \ref{sec:dis} we discuss the results of the combined X-ray spectra and MHD modeling of Wd2 to search for a possible non-thermal component which can be expected given the detected gamma rays from Wd2.  

\section{Multiwavelength view on RCW 49}\label{sec:multi}

Radio observations of Wd2 and the HII region RCW 49 with the Molonglo Observatory Synthesis Telescope (MOST)  and Australia Telescope Compact Array (ATCA) in the frequency bands between 843 MHz and 9 GHz were presented by \citet{radio_WU97} and \citet{Wd2_radio13}. Observations indicated that the bulk of the emission has flat spectral indices which are consistent with the emission of the optically thin thermal plasma. 
With the high resolution radio imaging \citet{radio_WU97} discovered two wind-blown shells in the core region of RCW 49. They associated the massive shell of $\sim$ $2.2^\prime$ thickness with the Wd2 cluster, while the second shell surrounds the massive star WR20b, which has an offset  $\sim 3.5^\prime$ from the center of Wd2.  The number density of electrons $n_e \approx 260$ \cmc was derived from the detected radio flux density in the massive shell. Then the estimated mass of the ionized gas in the Wd2 shell should be well above 100 \Msun. 
However, recent high resolution studies of the gas and dust in Wd2 surroundings with SOFIA and APEX telescopes \citep{Wd2_shell_SOFIA21}, using also the archival multiwavelength data, instead of the two radio shells revealed a single well-defined half-elliptical gas shell with the axes between $100^{\prime\prime}$ and $300^{\prime\prime}$. At the assumed distance of $\sim$ 4.2 kpc to Wd2  the sizes of the half-elliptical structure are about 2 - 6 pc with the thickness of $\sim$ 1 pc. The total mass of the gas  $\sim 10^4$ \Msun\ is moving with the velocity of $\sim$ 13 \kms\ to the observer. An estimated kinetic energy of the shell  
-- 4$\times 10^{49}$ erg --  can be provided by OB stars' winds over the lifetime of the cluster. Powerful WR 20a,b stars may also substantially contribute to the momentum and energy of the shell over 0.2-0.3 Myr time. 
 
\citet{Wd2_radio13} found some hints on the presence of the non-thermal radio emission with the spectral index values steeper than $-0.1$. The emission may be radiated either by relativistic electrons accelerated by colliding winds or by the secondary pairs, produced in the inelastic hadronic collisions of cosmic rays with the ambient gas. A diffuse emission with the flux density of about a few mJy (at most) was attributed to the vicinity of the Wd2 cluster. The apparent ridge with the steep spectra of the radio emission located to the south from Wd2 extends for about $7^\prime$ from NE to SW direction. It was pointed out by \citet{Wd2_radio13} that a search for the radio polarization in this region to establish firmly the non-thermal nature of the emission is challenging due to the high density of the thermal plasma providing a strong Faraday rotation. 

Westerlund 2 is a bright galactic gamma-ray source, detected by { \it Fermi} at GeV energies \citep{Ackerman17} and H.E.S.S. at TeV energies \citep{Wd2_Ahar11, Wd2_Mestre21}. Its gamma-ray spectrum has a spectral break on $\sim$TeV. This break may be explained in the frame of the kinetic model of particle acceleration and propagation in a highly turbulent plasma with shocks \citep{BK22}, assuming that the emission is of hadronic origin, i.e., due to the decay of neutral pions produced in $p-p$ collisions. The gamma-ray  spectrum of Wd2 continues at least up to $\sim$ 10 TeV, which implies that protons in a cluster must be accelerated up to $\sim$ 100 TeV energies.

A study of stellar and gas kinematics in Wd2 by \citet{Zeidler21} provided the estimated dynamical mass of the cluster $\sim 10^5$ \Msun which exceeds its photometric mass. Estimations of the distance to Wd2 cluster are not yet very certain  \citep[see e.g.][for a recent discussion]{Wd2_dist_18,Wd2_shell_SOFIA21} while the current data are in agreement with the distance of $\gsim$ 4 kpc derived by \citet{Wd2_dist13a} and \citet{Wd2_dist15}. The upper limit to the Wd2 age of 3 Myrs is consistent with the assumption that no supernovae have occurred yet in the cluster.

\section{Observations and data analysis}
\label{sec:obs}
\subsection{{\bfseries \itshape SRG}/ART-XC observations}

\begin{figure*}
	\includegraphics[width=\textwidth]{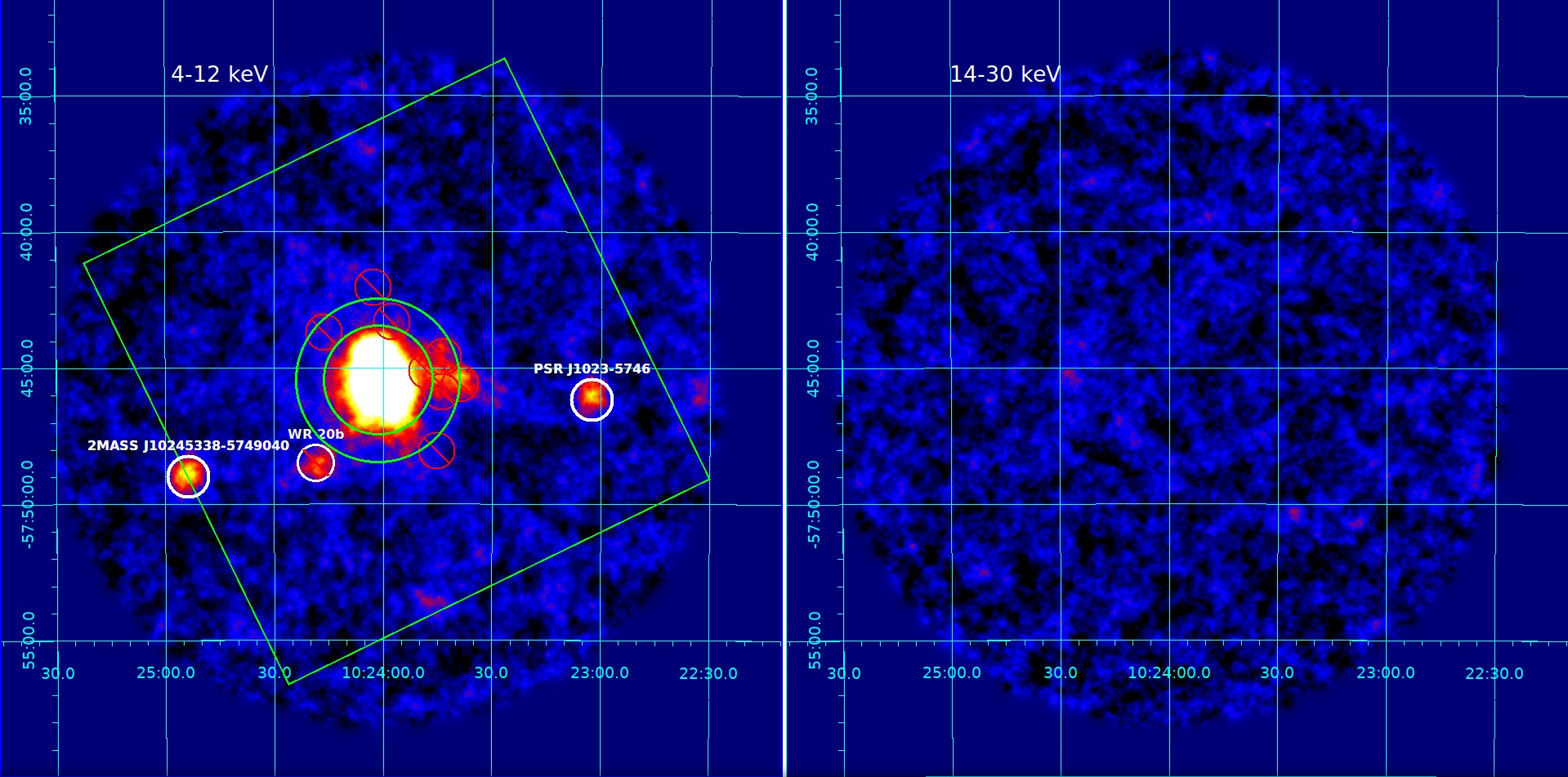}
    \caption{\sart\ map of Westerlund 2. The studied annulus region (see \S \ref{sec:subspec}) is shown in green together with the \chandra\ FOV (box region). The excluded regions are shown in red.
    Pulsar PSR J1023-5746, Wolf-Rayet star WR20b and young stellar object candidate 2MASS J10245338-5749040 are shown in white.}
    \label{fig:wd2-art}
\end{figure*}

The Mikhail Pavlinsky Astronomical Roentgen Telescope – X-ray Concentrator (ART-XC) \citep{ARTXC21} aboard the \srg\ orbital observatory \citep{SRG21} is a grazing-incidence-focusing X-ray telescope. The telescope includes seven modules sensitive in the 4–30 keV energy range. Each module is equipped with a CdTe double-sided strip detector with 48 $\times$ 48 pixels and a pixel size of approximately 45$^{\prime\prime}$, which makes the module’s field-of-view equal to 36$^{\prime}$. The \srg\ observatory is operating in the vicinity of the Lagrange Point L2 at a distance of $\sim1.5$ million kilometers from the Earth. This means that the detector background is determined by cosmic rays producing  $\sim 5\times 10^{-5}$ counts s$^{-1}$ keV$^{-1}$ arcmin$^{-2}$ in one telescope module. The detector background is not constant and depends on the solar activity. This makes observations of faint extended sources such as Wd2 especially challenging. 

The \art\ telescope observed Wd2 in June and July of 2022 (Fig. \ref{fig:wd2-art}), reaching a total exposure of about 320 ks. 
The ART-XC map in the 4-12 keV energy range explicitly shows the central part of the cluster with three other bright sources:  pulsar PSR\,J1023-5746 in the west, Wolf-Rayet WR\,20b
in the cluster neighborhood and a young stellar object candidate 2MASS\,J10245338-5749040 in the east (Fig.~\ref{fig:wd2-art}). Note that two other young stellar object candidates are located
within $30^{\prime\prime}$ circle from 2MASS\,J10245338-5749040. In the 14-30 keV energy range it is possible only to obtain an upper limit for the flux from the $2^\prime-3^\prime$ annulus region (our region of interest -- it is described in \S \ref{sec:subspec}), which is $1.4\times10^{-12}$ erg~s$^{-1}$~cm$^{-2}$ at the $3\sigma$ confidence level.

In order to extract correctly the extended emission, it is necessary to account for the particle background and exclude the cluster point sources. Unfortunately, the star cluster is rather dense and excluding all point sources from the studied region with the ART-XC point spread function (PSF, \citealt{ARTXC21}) is impossible. So the list of regions that are mostly contaminated with point sources was formed based on the data of \chandra\  observatory. These regions were excluded from the analysis. The particle background is not constant across the field-of-view. It is produced by the interaction of cosmic rays with the telescope structure, which results in secondary X-rays hitting the detector. The particle background is typically higher near the detector edges. In order to account for it, the model background was collected during observations with no sources in the field-of-view (survey and scan spacecraft operational modes). This model background was then normalized to the background conditions of a particular observation and accounted for the following data reduction. 

Furthermore, an additional analysis has shown that, due to a relatively wide ART-XC PSF, a fraction of the flux from the central region is added to the extended emission in the $2^\prime-3^\prime$ annulus. In order to estimate this fraction we have convolved the 4-8 keV \chandra\ image with the PSF of ART-XC. This has shown that the Wd2 central core adds $\simeq5$\% of its flux to the $2^\prime-3^\prime$ annulus. Thus, to obtain the net flux of the extended emission in the region of interest we subtracted 5\% of the central region spectrum from the spectrum of the $2^\prime-3^\prime$ annulus. 

\subsection{{\bfseries \itshape Chandra} X-ray Observatory data}

To distinguish the diffuse X-ray emission from the numerous point sources in Wd2 stellar cluster we used the existing archival observations of Wd2 region made with the \chandra\ observatory \citep{Chandra02} providing the sensitive observations with sub-arcsecond angular resolution up to 8 keV photon energy. Namely, the observations performed in 2018 with ObsIDs 20133, 20134, 21842, 21847, 21848 ({PI: Laura Lopez}) and
total exposure of $2.6\times10^5$ s were analysed together with ART-XC data at higher energy X-rays. All the \chandra\ observations were done in the ACIS-I 
configuration and the same aimpoint, offset and rotation angle. Observations were
reprocessed with CIAO  \citep{2006SPIE.6270E..1VF} ver. 4.14 and CALDB 4.9.3. The flare correction was done for ObsID 20133. 
All spectra were obtained with {\it specextract} and {\it combine\_spectra} scripts
with grade=0 filter applied that increases the  quality of the processed data for these observations. The   XSPEC%
\footnote{http://heasarc.gsfc.nasa.gov/xanadu/xspec/%
} package  v. 12.12.0 was used for the spectral analysis \citep{xspec_1996ASPC..101...17A}.
All spectra were grouped with not less then 100 counts in a bin.

\section{Results}
\label{sec:an}
\subsection{Spectral analysis}
\label{sec:subspec}
To study the diffuse emission of the cluster we needed to exclude all point sources from the \chandra\ Source Catalog\footnote{http://cxc.cfa.harvard.edu/csc/} (CSC) v2 from the region of analysis. The total number of $\sim900$ point sources located at the \chandra\
field of view (FOV) were excluded as circle regions with $5^{\prime\prime}$ radius. Most of them are located at the
center of the cluster, overlapping with each other. The significant part of the central
cluster area also falls at the inter chip area of the \chandra\ detector. Because of all this the $2^\prime-3^\prime$ annulus region with center coinciding with the cluster center was chosen for the analysis of the cluster diffuse emission.

\begin{figure}
	\includegraphics[width=\columnwidth]{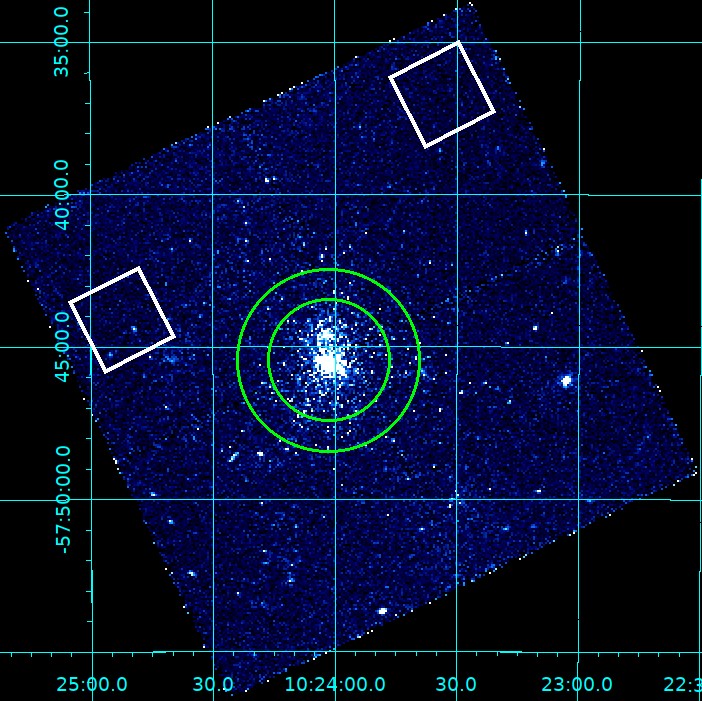} \\
    \includegraphics[width=\columnwidth]{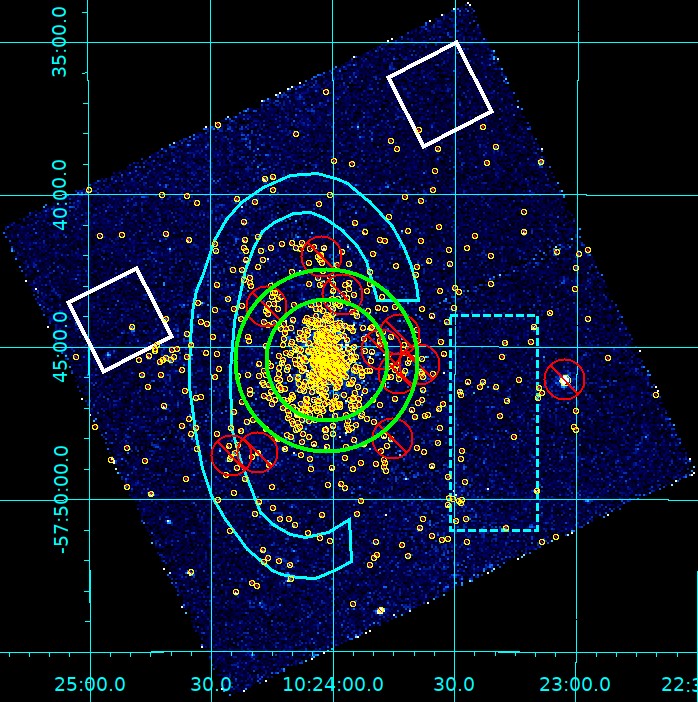}
    \caption{\chandra\ 2-7 keV energy map of Wd2. $2^\prime-3^\prime$ annulus region  used for 
    spectrum analysis is green, background region is white. At the lower panel the excluded point source regions from the
    CSC v.2 are yellow, additionally excluded groups of bright sources are red, the arc region is cyan and the west box region is dashed cyan. }
    \label{fig:Chandra}
\end{figure}

We focused on  a joint spectral analysis of ART-XC and \chandra\ data. ART-XC point spread function (PSF) is $\sim30^{\prime\prime}$ and exclusion of all point sources from the $2^\prime-3^\prime$ annulus region as circles with such radius would too strongly reduce the area suitable for analysis. So only few regions with $40^{\prime\prime}$ -- the most bright groups of point sources -- were excluded 
from the analysis of ART-XC data. The same regions were  excluded from the analysis of \chandra\ data. 
The 2-7 keV \chandra\ map with marked regions of data analysis is shown in Fig. \ref{fig:Chandra}.
However, the additional excluded point sources from the \chandra\ data, different PSFs and sensitivities of both detectors make necessary to renormalize 
ART-XC and \chandra\ model fluxes in the spectrum analysis. This renormalization was done by addition of constant multiplicative model component $A$
for spectral models, which value for \chandra\ equals 1  and for ART-XC is found during fitting. That is, if \chandra\ spectrum
model is $J(E)$, then ART-XC spectrum model is assumed to be $A\cdot J(E)$. 

\begin{figure}
	\includegraphics[width=\columnwidth, viewport= 50 65 720 520, clip=no]{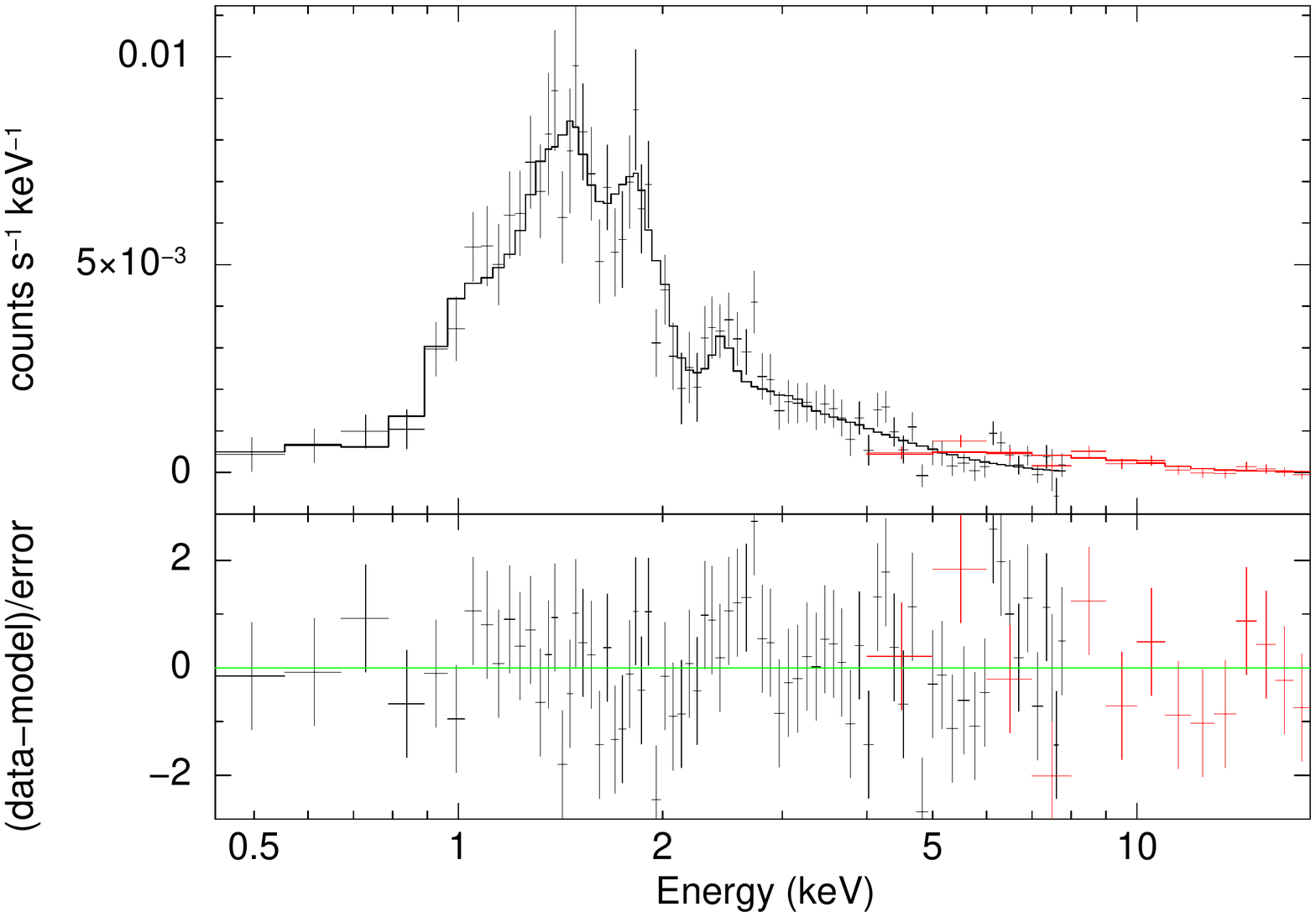}
    \caption{Upper panel: spectrum data and model $tbabs \times apec+tbabs \times (apec+pow)$ of the $2^\prime-3^\prime$ annulus region. Lower panel: residuals. Black and red colors show \chandra\ and ART-XC spectra respectively.}
    \label{fig:sp1}
\end{figure}

Likewise in  \citet{Townsley2005} and \citet{Wd2_Suzaku09} the
multicomponent spectral models were used with (i) three thermal emission components (hereafter 3T model) and (ii) two thermal components with one power-law component (hereafter 2TP). APEC spectral model 
was used for plasma thermal emission of collisionally ionized diffuse gas simulation using atomic data from the AtomDB%
\footnote{http://www.atomdb.org/%
} database. APEC spectral data
v.3.0.9 and eigenfunction data v.3.0.4  were used in XSPEC for the simulations. A TBABS interstellar absorption model with corresponding abundances \citep{Wilms_2000ApJ...542..914W} was used to calculate  the interstellar absorption.
Following \citet{Wd2_Townsley19}, the temperature of the softest thermal emission component was frozen at $T_1=0.1$ keV (this value was obtained earlier by \citealt{Townsley2005} and   \citealt{Wd2_Suzaku09}). The column density for this component is poorly constrained and is of the order of $10^{21}$ cm$^{-2}$ (\citealt{Townsley2005}, see also Table 5 in \citealt{Wd2_Townsley19}, where the {\it pshock} model was used). Much larger values of $N_{\mathrm{H}}\simeq(2-3)\times10^{22}$ cm$^{-2}$ are  expected for the Wd2 region from the extinction magnitude analysis  \citep[see e.g.][]{Wd2_dist13a,Wd2_shell_SOFIA21}. Therefore, we assume that this soft component is likely related to some  foreground emission and it is not associated with the Wd2 region which is the subject of our paper.
The fitting results for both models are listed in two first columns of Table \ref{tab:tab_ann1_1} for \chandra\ data fit only and Table \ref{tab:tab_ann2_1} for the joint \chandra\ +ART-XC fit. Fit parameters correspond very well within 90\% confidence interval with the results of \citet{Townsley2005} and \citet{Wd2_Suzaku09}.   The joint spectrum for a model with 2 thermal+1 power-law components is shown in Fig. \ref{fig:sp1}. Note that the \chandra\ spectrum has a little bump at energies close to 6 keV. Adding the 6.4 keV Fe line to analysis allows only a slight improvement of the fit quality.

\renewcommand{\arraystretch}{1.375}
\begin{table*}
\caption{\label{tab:tab_ann1_1} \chandra\ spectrum models of the Wd2 emission from $2^\prime-3^\prime$ annulus region. }
\begin{center}
\begin{tabular}{lccc}
\hline 
 model   & 3T: $tbabs_1 \times apec_1~+$   & 2TP: $tbabs_1 \times apec_1~+$   & TMP: $tbabs_1 \times apec_1~+$  \\
 parameters   & $tbabs_2 \times (apec_2+apec_3)$   & $tbabs_2 \times (apec_2+pow)$   & $tbabs_2 \times (mhd\_th+pow)$  \\
\hline
$N_{\mathrm{H}}^{\mathrm{I}}$\hspace{5pt}($10^{22}$ cm$^{-2}$)  & $0.1_{-0.1}^{+0.7}$  
 & $0.13_{-0.13}^{+0.67}$   & $0.15^{*}$  \\

$T_1$\hspace{9pt}(keV, fixed)  & $0.1$  & $0.1$  & $0.1$  \\

$F_{T_1}^{**}$\hspace{5pt}(erg cm$^{-2}$ s$^{-1}$)   & $1.0 ^{+11.6}_{-0.9} \times 10^{-13}$ &  $1.6 ^{+14.4}_{-1.5} \times 10^{-13} $ & ($2.4 \pm {2.1}) \times 10^{-13}$\\

$N^{\mathrm{II}}_{\mathrm{H}}$\hspace{5pt}($10^{22}$ cm$^{-2}$)  & $1.5 \pm {0.3}$  & $1.6^{+0.25}_{-0.35}$  & $0.9 \pm 0.2$  \\

$T_2$\hspace{9pt}(keV)  & $0.85^{+0.25}_{-0.35}$  & $0.9^{+0.3}_{-0.2}$   & --   \\

$T_3$\hspace{9pt}(keV)  & $3.8^{+8.2}_{-1.3}$   & --   & --   \\

$\Gamma$\hspace{12pt}(power-law index)  & --  & $1.9^{+0.5}_{-0.7}$  & $2.4 \pm 0.2$ \\

$F_{T_2}^{**}$\hspace{5pt}(erg cm$^{-2}$ s$^{-1}$)  & $4.0^{+2.3}_{-2.4} \times 10^{-13}$   & $5.0 ^{+2.9}_{-3.0} \times 10^{-13} $ & --  \\

$F_{T_3}^{**}$\hspace{5pt}(erg cm$^{-2}$ s$^{-1}$)  & $2.0^{+0.5}_{-0.7} \times 10^{-13}$   & --   & -- \\

$F_{mhd\_th}^{**}$\hspace{5pt}(erg cm$^{-2}$ s$^{-1}$)  & --  & --  & $1.3 \times 10^{-13}$  \\

$F_{pow}^{**}$\hspace{15pt}(erg cm$^{-2}$ s$^{-1}$)  & --  & $2.0^{+1.2}_{-0.7} \times 10^{-13}$  & $3.6 ^{+0.9}_{-0.4} \times 10^{-13}$  \\

$dof$  & 69  & 69  & 72  \\

$\chi^{2}/dof$   & 1.11   & 1.16   & 1.29  \\
\hline 
\multicolumn{4}{l}
{\begin{minipage}
{0.8\textwidth}
$^{*}$ Here $N_{\mathrm{H}}^{\mathrm{I}}$ is fixed. \\ 
$^{**}$ Unabsorbed flux (erg cm$^{-2}$ s$^{-1}$) in the 0.5-8.0 keV energy range.  \\
All errors are listed with 90\% confidence level. 
\end{minipage}}
\end{tabular}
\end{center}
\end{table*}

\begin{table*}
\caption{\label{tab:tab_ann2_1} Combined \chandra\ and ART-XC spectrum models of the Wd2 emission from $2^\prime-3^\prime$ annulus region. }
\begin{center}
\begin{tabular}{lccc}
\hline 
 model   & 3T: $tbabs_1 \times apec_1~+$   & 2TP: $tbabs_1 \times apec_1~+$   & TMP: $tbabs_1 \times apec_1~+$  \\
 parameters   & $tbabs_2 \times (apec_2+apec_3)$   & $tbabs_2 \times (apec_2+pow)$   & $tbabs_2 \times (mhd\_th+pow)$  \\
\hline 
$N_{\mathrm{H}}^{\mathrm{I}}$\hspace{5pt}($10^{22}$ cm$^{-2}$)  & $0.1_{-0.1}^{+0.7}$  &  $0.18_{-0.18}^{+0.75}$ & $0.15^{*}$  \\

$T_1$\hspace{9pt}(keV, fixed)   & $0.1$   & $0.1$   & $0.1$  \\

$F_{T_1}^{**}$\hspace{5pt}(erg cm$^{-2}$ s$^{-1}$)   & $1.0^{+9.0}_{-0.9} \times 10^{-13}$   & $1.8^{+16.2}_{-1.6} \times 10^{-13} $   & ($1.5 \pm 1.3) \times 10^{-13}$ \\

$N^{\mathrm{II}}_{\mathrm{H}}$\hspace{5pt}($10^{22}$ cm$^{-2}$)   & $1.5^{+0.3}_{-0.2}$ 
  & $1.5 \pm 0.3$   & $0.9 \pm 0.2$  \\

$T_2$\hspace{9pt}(keV)   & $0.9^{+0.2}_{-0.3}$   & $0.9^{+0.4}_{-0.2}$   & --  \\

$T_3$\hspace{9pt}(keV)   & $4.7^{+13.3}_{-1.9}$   & --   & --  \\

$\Gamma$\hspace{12pt}(power-law index)  & --  & $2.1^{+0.3}_{-0.4}$  & $2.4 \pm 0.2$  \\

$F_{T_2}^{**}$\hspace{5pt}(erg cm$^{-2}$ s$^{-1}$)   & $5.0^{+2.4}_{-2.4} \times 10^{-13}$   & $4.0^{+2.3}_{-2.4} \times 10^{-13} $   & --  \\

$F_{T_3}^{**}$\hspace{5pt}(erg cm$^{-2}$ s$^{-1}$)   & $1.8 ^{+0.5}_{-0.7} \times 10^{-13}$   & --   & --  \\

$F_{mhd\_th}^{**}$\hspace{5pt}(erg cm$^{-2}$ s$^{-1}$)   & --   & --   & $1.3 \times 10^{-13}$  \\

$F_{pow}^{**}$\hspace{15pt}(erg cm$^{-2}$ s$^{-1}$)   & --   & $2.2^{+1.0}_{-0.7} \times 10^{-13}$   & $3.3 ^{+0.6}_{-0.4} \times 10^{-13}$  \\

$A^{***}$  & $1.85^{+0.85}_{-0.65}$  & $1.95^{+0.85}_{-0.60}$  & $2.1^{+0.7}_{-0.5}$  \\

$dof$   & 82   & 82   & 85  \\

$\chi^{2}/dof$   & 1.16   & 1.15   & 1.24  \\
\hline 
\multicolumn{4}{l}
{\begin{minipage}
{0.8\textwidth} 
$^{*}$ Here $N_{\mathrm{H}}^{\mathrm{I}}$ is fixed. \\
$^{**}$ Unabsorbed flux (erg cm$^{-2}$ s$^{-1}$) in the 0.5-8.0 keV energy range. \\
$^{***}$ Spectral model renormalization coefficient. If \chandra\ spectrum model is $J(E)$ then ART-XC spectrum model is $A\cdot J(E)$.  \\
All errors are listed with 90\% confidence level.
\end{minipage}}
\end{tabular}
\end{center}
\end{table*}

\subsection{Magnetohydrodynamic modeling of Westerlund 2 region}

Both 3T and 2TP models provide adequate fitting for \chandra\ data as well as for joint observations. However, for 3T model the joint fit requires high temperature of the third thermal component ($\sim 5$ keV), which may be difficult to reach in the cluster. To see which plasma temperatures can be obtained in the cluster core and, therefore, to distinguish between purely thermal and thermal+power-law models we performed 3D MHD modeling of an inner structure of a cluster with the help of the PLUTO numerical code \citep[][]{Mig07}. The same modeling, but representing Westerlund 1, is described in detail in \citet{Bad22}. For this paper we adjusted the size and stellar content of our model cluster to reproduce the main characteristics of Wd2. We focused on the massive stars with fast winds and high mass-loss as they mostly determine the kinetic energy release in the cluster. So, we put 20 massive stars with the total mechanical power $\dot{E}_{\mathrm{cl}}=2 \times 10^{38}$\ergs, as expected in Wd2, into the 2.4 pc radius, which for distance to Wd2 of $\sim$ 4.4 kpc corresponds to $2^\prime$. The full size of simulation box was 7.3 pc, so we could explore the region, corresponding to $2^\prime-3^\prime$ annulus.  
\begin{figure}
	\includegraphics[width=\columnwidth,viewport= 5 5 700 600, clip=true]{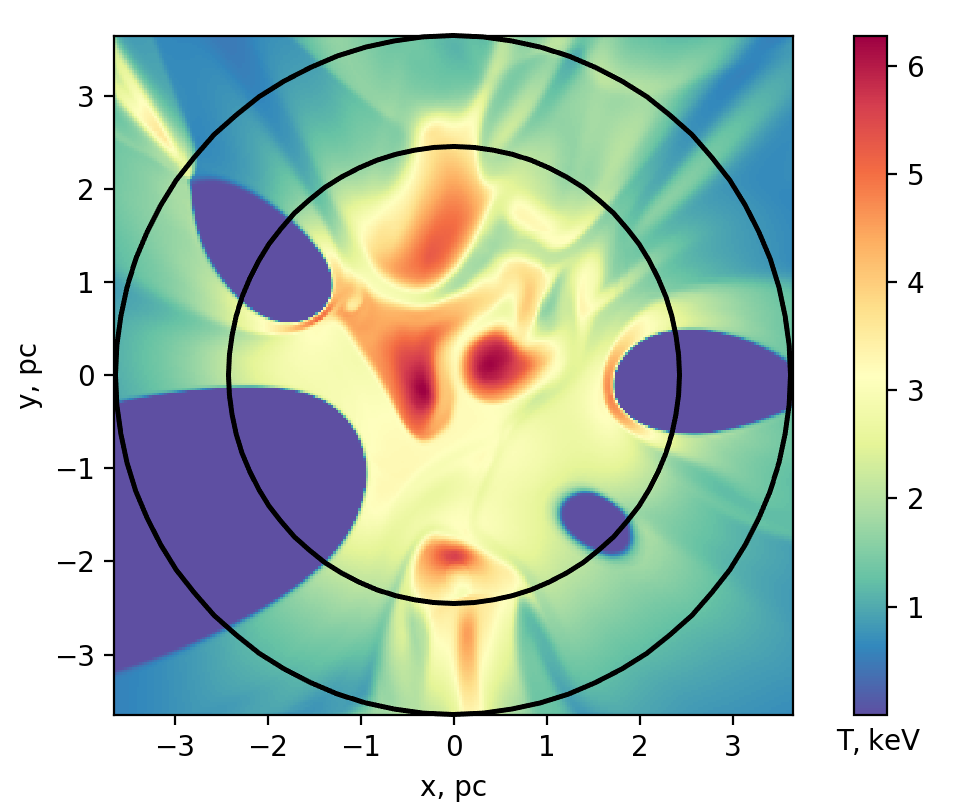}
    \caption{Temperature map of the central plane of the cluster simulated with the single-fluid 3D MHD model. Black circles indicate the positions of the $2^\prime-3^\prime$ annulus region borders.}
    \label{fig:tpluto}
\end{figure}

\begin{figure}
	\includegraphics[width=\columnwidth,viewport= 50 65 720 520, clip=true]{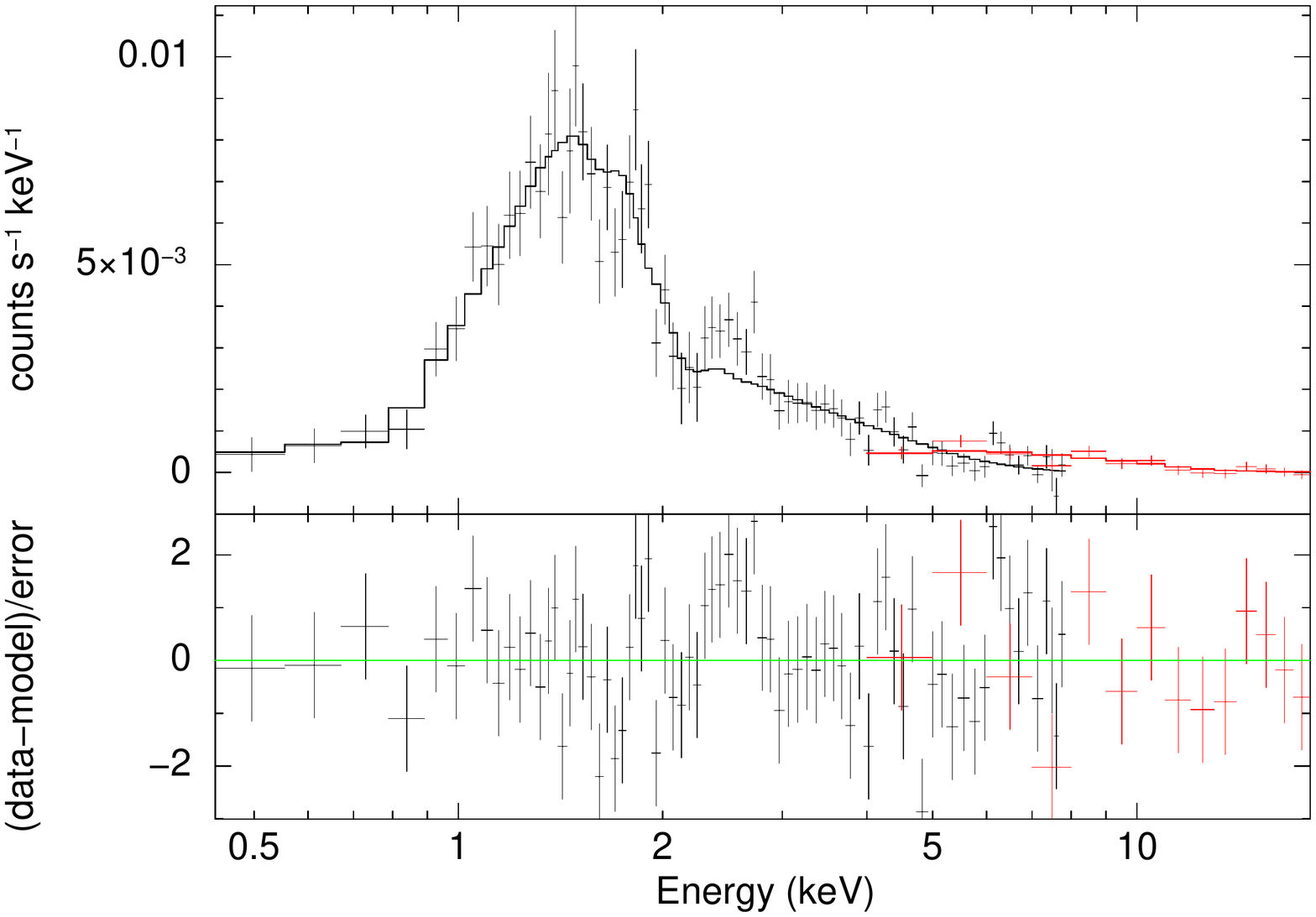}
    \caption{Upper panel: spectrum data and model $tbabs \times apec+tbabs \times (mhd\_th+pow)$ of the $2^\prime-3^\prime$ annulus region.  Lower panel: residuals. Black and red colors show \chandra\ and ART-XC spectra respectively.}
    \label{fig:sp2}
\end{figure}

The modeling allows mapping plasma density, temperature and the magnetic field structure in the vicinity of the cluster core. Interactions of colliding fast winds of massive stars lead to plasma heating and turbulent magnetic field amplification in the cluster. In Fig. \ref{fig:tpluto} the temperature map of the central plane of the cluster is shown, with the marked positions of $2^\prime$ and $3^\prime$ circles. One can see that in the $2^\prime-3^\prime$ annulus there are both cold regions of massive stellar winds and regions with heated up to $\sim$6 keV plasma.

In the single-fluid MHD model the electron temperature $T_e$, which determines the emission of optically thin plasma, can be estimated in the following way. In the case of the local electron-ion collisional equilibration $T_e$ can reach $T/2$, where $T$ is the single-fluid plasma temperature. It can also be lower, which depends on the thermal conduction in the cluster. The recipes of estimating the electron temperatures in the cluster are discussed in \citet{Bad22}, while here for simplicity we assume $T_e \approx T/2$.  Having determined electron temperature and plasma density in each simulation bin, we managed to  calculate the thermal X-ray spectrum from the $2^\prime-3^\prime$ annulus,  using APEC spectral model and integrating the fluxes from all bins in this region. Using the functionality of PyXspec, the resulting spectrum was then added as an independent model (referred as $mhd\_th$) to XSPEC fitting procedure, described in \S \ref{sec:subspec}. Note that $mhd\_th$ has no variable parameters: its spectral form and normalization are fixed and determined by plasma temperatures obtained in MHD modeling and energy release of Wd2. 

If the X-ray emission of Wd2 was purely thermal, one would expect $tbabs \times apec+tbabs \times mhd\_th$ model to provide a good fit to the data. However, this is not happening: {$\chi^2/dof>2$} for such fit, because the obtained in MHD simulation electron temperatures are well below 4-5 keV needed for the thermal explanation of data (see the first columns of Tables \ref{tab:tab_ann1_1},\ref{tab:tab_ann2_1}). It is worth mentioning that such temperatures are not reached even in \citet{Bad22} modeling of a cluster with a 5 times higher energy release than Wd2. Thus, according to MHD modeling, it is unlikely that all of Wd2 X-ray emission is thermal.

To verify the hypothesis of a presence of a non-thermal emission we performed a spectral fitting with one thermal component with a fixed $T_1=0.1$ keV, one thermal component obtained from MHD modeling and a power-law component 
(hereafter TMP model).  This fit is rather insensitive to the $N_{\mathrm{H}}^{\mathrm{I}}$ parameter value
so it was fixed equal to $0.15 \times 10^{22}$ cm$^{-2}$ -- close to the best fit values obtained for other spectral models. The results of the fits for \chandra\ data only and for \chandra\ +ART-XC data are shown in third columns of Tables \ref{tab:tab_ann1_1}, \ref{tab:tab_ann2_1}. The fits show acceptable quality, despite the fact that $mhd\_th$ component cannot be adjusted in any way. The indices of the power-law component in TMP model fits coincide within errors with the values obtained in 2TP model fits, as well as in fits by \citet{Townsley2005} and \citet{Wd2_Suzaku09}. One also can see that the joint fit is slightly better than the \chandra\-only. The corresponding joint spectrum is shown in Fig. \ref{fig:sp2}. 

\subsection{{\bfseries \itshape Chandra} spectrum of the arc shell region}

To search for the non-thermal emission component from particles which were accelerated in Wd2 and then left the cluster, we selected for the additional  spectral analysis the arc region (see the cyan region in Fig. \ref{fig:Chandra}). The selected  region coincides with the expanding shell feature discovered in [C II] and  ${^{12}}$CO line observations by \citet{Wd2_shell_SOFIA21} (see their Fig. 7). The arc has a dense photo-dissociation region with the gas density of about 4,000 \cmc. The massive arc shell is likely accelerated by the momentum flux emitted by WR 20a,b stars over the last 0.5 Myrs \citep{Wd2_shell_SOFIA21}. The magnetic field magnitudes of about 50 $\mu$G can be expected in the dense regions of the shell according to the magnetic field measurements in galactic molecular clouds reviewed by \citet{1999ApJ...520..706C,MF_Crutcher12}. Such fields may provide favorable conditions for synchrotron X-ray emission as we'll discuss below in \S\ref{synchr}. ART-XC flux from the shell region is too low to be detected, so, to study the possible non-thermal component of the arc shell we constructed  \chandra\  spectrum of the arc shell.
By means of grouping with $\gsim$300 counts in a bin there is a possibility to check if a model with the non-thermal component, which we got in the annulus region, can provide an adequate fit. To do so we fixed the power-law photon index to be equal to 2, i.e. close to that in 2TP and TMP fits of annulus region in the Table \ref{tab:tab_ann2_1}.  
The three-component spectral model $tbabs_1 \times apec_1+tbabs_2 \times apec_2+tbabs_3 \times pow$ with fixed $N^{\mathrm{I}}_{\mathrm{H}}=0.15\times10^{22}$cm$^{-2}$,  
$T_1=0.1$~keV and power-law index $\Gamma=2$ approximate the data with $\chi^{2}/dof=1.1$ ($dof=43$). The fit parameters are $N^{\mathrm{II}}_{\mathrm{H}}=1.3_{-0.8}^{+0.5}\times10^{22}$ cm$^{-2}$,
$T_2=1.3^{+0.5}_{-0.3}$~keV, $N^{\mathrm{III}}_{\mathrm{H}}=5.6_{-2.8}^{+4.9}\times10^{22}$ cm$^{-2}$ (all errors are listed with 90\% confidence).
The unabsorbed flux in the power-law component from the arc is about $3.3_{1.1}^{+1.2}\times 10^{-13}\, \efl$.  The hot plasma with $T_2 \sim$ keV and $N^{\mathrm{II}}_{\mathrm{H}}$ close to that was found in the annulus region can be produced by the collision of Wd2 cluster wind of velocity $\gsim 1,000$ \kms with 
the dense shell revealed 
by \citet{Wd2_shell_SOFIA21}. The non-thermal synchrotron radiation is likely produced by multi-TeV electrons in very dense regions of the shell with high magnetic field. This  results in the higher column density of the power-law component.   

We also analysed \chandra\ diffuse spectrum in a box region of a similar size located to the west from the Wd2 core (the dashed rectangle in Fig. \ref{fig:Chandra}) and found no significant evidence for a hard power-law component which is detected from the arc shell region. This could be understood if the power-law component in the arc shell is of a synchrotron origin since rather high magnetic field magnitudes are required to get synchrotron X-rays. These high fields can be associated with the dense gas observed by \citet{Wd2_shell_SOFIA21} in the arc. In the western region no that dense gas is apparent.

\section{Discussion}
\label{sec:dis}
We showed above that the combined ART-XC and \chandra\ 0.4-20 keV data taken from the $2^\prime-3^\prime$ annulus just outside the core of Wd2 can be fitted equally well with both models: (i) three temperature model and (ii) two temperature plus non-thermal power-law emission -- the result consistent with the fits previously obtained with just \chandra\ \citep{Townsley2005,Wd2_Townsley19} and {\sl Suzaku} \citep{Wd2_Suzaku09} data in different regions of Wd2. The three temperature model requires a presence of rather hot electrons with T$_e \sim 5$ keV in the annulus. Modeling of the cluster structure discussed above does not allow getting that high electron temperatures, so the purely thermal emission can hardly be explained. This suggests considering in more detail the non-thermal origin of X-ray radiation above a few keV.     

Non-thermal bremsstrahlung origin of the power-law X-ray component is not really likely given the very low photon yield of the Coulomb collision process \citep[see e.g.][]{Petrosian01}, which implies that to get the X-ray luminosity of $\sim 10^{32}$ \ergs from the annulus or the arc shell one must dissipate there the power above 10$^{37}$ \ergs. Moreover, the penetration of $\gsim$ keV regime particles into the dense arc shell is not possible. Therefore, one may consider the non-thermal X-ray emission component to be provided by relativistic electrons of very high energies (synchrotron radiation) or mildly relativistic electrons (inverse Compton radiation).

\subsection{Synchrotron X-ray radiation model}\label{synchr} 
The synchrotron X-ray emission from the cluster can be modeled in detail with the help of a kinetic model of particle acceleration in Wd2, applied to high-energy gamma-ray emission \citep{BK22}, and MHD modeling of magnetic fields in Wd2. The MHD-simulated structure of magnetic fields in a cluster with colliding powerful stellar winds demonstrates highly intermittent character 
\citep{Bad22}.  In our MHD model of Wd2 we examined the magnetic field magnitudes near the cluster core: most of the volume is filled with magnetic fields below 10 $\mu$G, but there are regions with high magnitudes > 50 $\mu$G which are filling a few percent of the core volume. These regions are dominating the magnetic energy of the system.        
The intermittent structure of the magnetic field is particularly important for modeling the electron acceleration and radiation in the system. Very high energy electrons can be accelerated in regions with low magnetic fields, filling most of the volume (Compton losses on the optical photons are reduced for them in the Klein-Nishina regime), while synchrotron X-ray radiation of the accelerated electrons can be produced in the domains with high magnetic field, filling a few percent of the volume.    

The spectrum of protons, accelerated by multiple shocks of colliding winds in a cluster, can be used to model the non-thermal electrons once the appropriate synchrotron-Compton energy losses are included. The maximal Lorentz factor $\gamma_e^{max}$ of electrons accelerated by diffusive Fermi shock acceleration can be estimated from the comparison of the acceleration time and the time of the electron energy losses within the accelerator. For example, for Bohm diffusion regime one can estimate $\gamma_e^{max}$ from a simple relation:      
\begin{equation}
\gamma_e^{max} \approx 5\times 10^4 u_8 \sqrt{\frac{B_{st}}{W_0}},   \label{gmax} 
\end{equation}
 where $u_8$ is the r.m.s. velocity of large scale turbulence with multiple wind shocks 
 measured in 1,000 \kms units, $B_{st}$ is the r.m.s. amplitude of magnetic fluctuations measured in Gauss, and  $W_0$ (measured in \ergcc) is the energy density of the magnetic field and radiation corrected for the Klein-Nishina effect. The optical radiation field photons are scattered by electrons of Lorentz factors $\gamma > 10^5$ (which are of most interest for synchrotron X-ray radiation) in the Klein-Nishina regime. Therefore, the contribution of the optical radiation energy density to $W_0$ is far less than full $W_{\rm opt}$ in this regime \citep[see e.g.][]{Moderski05}.
 In the magnetic field of amplitude $\lesssim$ 10 $\mu$G,  one can get from Eq. \ref{gmax} $\gamma_e^{max} \sim 10^8$. Being accelerated in the relatively low magnetic field domains, the electrons can radiate synchrotron photons above a few keV in the regions of higher magnetic fields in the cluster core and its surrounding dense shell.   

\begin{figure}
	\includegraphics[width=\columnwidth]{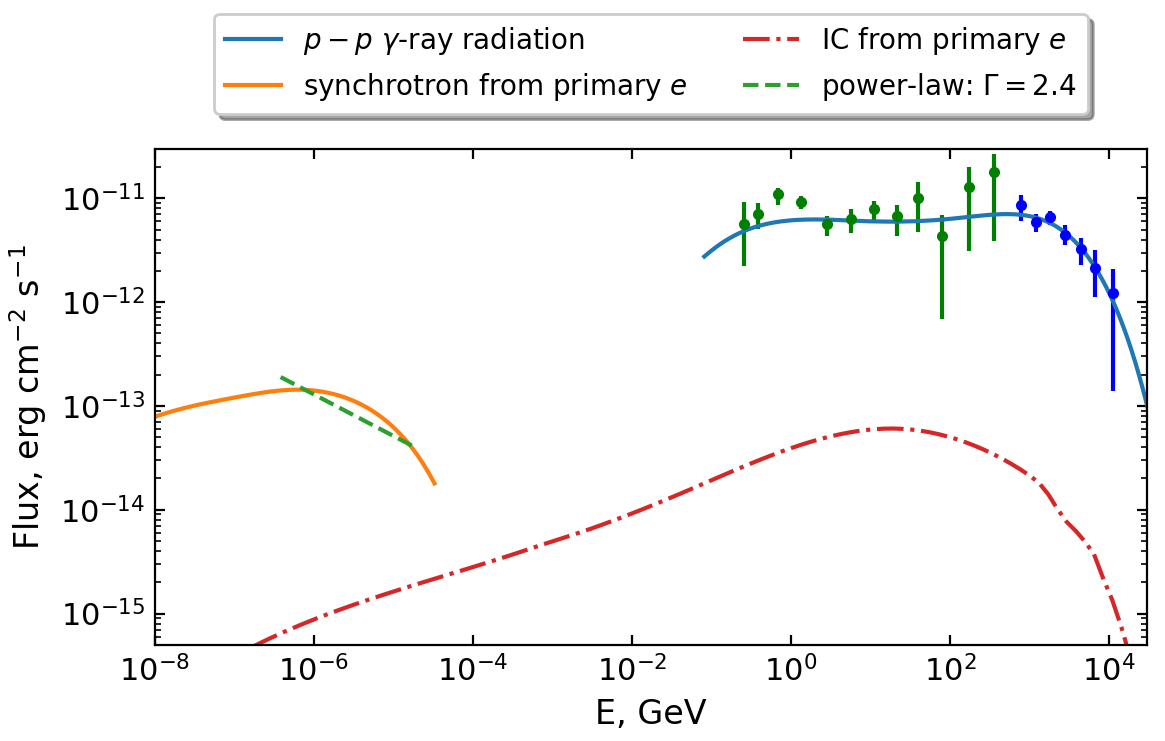}
    \caption{The results of the modeling of synchrotron (orange solid line) and inverse Compton (red dot-dashed line) emission in $2^\prime-3^\prime$ annulus from electrons and positrons directly accelerated in Wd2 core region. Gamma-ray data from the {\it Fermi} observatory (green dots, \citealt{Ackerman17}) and H.E.S.S. telescopes (blue dots, \citealt{Wd2_Ahar11,Wd2_Mestre21}) are shown as well as the model of the Wd2 gamma-ray radiation presented in \citet{BK22} (blue solid line). Green dashed line corresponds to the non-thermal component obtained in the XSPEC \chandra\  +ART-XC spectral fitting with the TMP model. }
    \label{fig:synch}
\end{figure}

 The electrons and positrons with energies of tens of TeV,  both primary (directly accelerated in Wd2) and secondary (pairs, provided by the decay of charged pions from the inelastic collisions of multi-TeV accelerated protons with dense cloud matter),  can emit keV regime X-ray synchrotron radiation. The presence of the very high energy protons accelerated in Wd2 is supported by the conclusion about the hadronic origin of TeV photons detected from Wd2 \citep{Wd2_Yang18} as well as from Westerlund 1 and some other clusters, emitting in gamma rays (see for a discussion \citet{Aha19}). 
 According to the MHD modeling of the core of Wd2, high magnetic fields with magnitudes > 50 $\mu$G and up to $\sim 300~ \mu$G are filling $\sim 3\%$ of a simulation box volume. This allowed us to estimate the expected synchrotron radiation from the annulus region, where we obtained the non-thermal component with the power-law index $\Gamma \sim 2.4$ and flux $\sim 3 \times 10^{-13}~\efl$ (see third column of Table \ref{tab:tab_ann2_1}). From our modeling of gamma-ray emission of the Wd2 cluster core with $p-p$ mechanism, presented in \cite{BK22}, we obtained the spectrum of protons which can fit {\it Fermi}  \citep{Ackerman17} and H.E.S.S. \citep{Wd2_Ahar11,Wd2_Mestre21} observations. The same model provided the spectrum of primary electrons, accelerated in the cluster.  The lepton spectrum repeats the shape of proton spectrum up to the maximal energy allowed by the synchrotron and Compton losses in the acceleration region. Within the kinetic model we obtained that in the low mean magnetic field $\lesssim 7~ \mu$G, which is filling most of the cluster volume, the maximal electron energies can be as high as $\sim 50$ TeV. The total energy in accelerated electrons can be a few percent of that in protons. To fit the synchrotron X-ray spectra  we chose the electron spectra normalization containing about 10\% of the energy in accelerated protons. Then we modeled the spectrum and flux of synchrotron emission of these electrons in $2^{\prime}-3^{\prime}$ annulus, taking into account the magnitude distribution of high (> 50 $\mu$G) magnetic fields in a cluster and their volume filling factor. The result is shown in Fig. \ref{fig:synch}. The model allows reproducing the maximal energy, normalization and, for most of the energies, the spectral shape of the non-thermal component obtained in spectral fitting. 
 
 One can see from Fig. \ref{fig:synch} that within the frame of the model the gamma-ray spectrum falls down rather fast after $\sim 10$ TeV. This limits the maximal energies of the secondary e$^\pm$ pairs, born from the decay of charged pions that accompanied the gamma-ray production. To make a significant contribution to the $\gsim$ keV-energy range in the annulus, the secondary pairs must radiate the synchrotron photons in the regions with magnetic field magnitudes well above hundreds of $\mu$G, which is not very likely.  However, it is uncertain yet if there is a cutoff in the gamma-ray spectrum at the energies higher than the last point of H.E.S.S. observations. Also, highly turbulent magnetic fields could extend somewhat the synchrotron photon spectra to allow the secondary lepton model. 
 Moreover, the topology and the volume filling factors of the high magnetic fields in the dense regions surrounding Wd2 reported by \citet{Wd2_shell_SOFIA21} are not yet known. The expected positive correlation of the magnetic field magnitude with the gas density allows establishing an upper limit of $\sim$ $10^{-12}\, \efl$ on the synchrotron radiation energy flux from the secondary  $e^{\pm}$ pairs produced simultaneously with the gamma rays from the whole region of the VHE gamma-ray source ($\sim 20$ pc).

\subsection{Inverse Compton radiation}
X-ray photons of keV energies in the vicinity of young massive star cluster can be produced
by the IC scattering of local radiation field photons by mildly relativistic electrons accelerated 
in the cluster \citep[see e.g.][]{Mun06,Wd2_Bedanarek97,IC_Wd207}.

The primary electron component in our model has a power-law spectrum with index close to 2 on energies $\lesssim$ TeV. We extrapolated this spectrum down to $\sim$ MeV energies in order to estimate the X-ray IC radiation in the annulus region (Fig.~\ref{fig:Chandra}). Note that we used the same population of primary electrons, which was discussed in \S\ref{synchr}. In the vicinity of the cluster core full of luminous stars the most energetic component of target photon field is the cluster optical radiation.
We assumed that it has temperature $T=10000-60000$ K. The exact value does not affect the result much in this case; we took $T=30000$ K for our illustration (Fig. \ref{fig:synch}). The energy density of the cluster photons in 3$^\prime$ radius was taken as  $U_\textrm{opt} = 600$~eV~cm$^{-3}$ according to the radiation pressure and Wd2 bolometric luminosity estimates given in Table 1 of  
\cite{Wd2_shell_SOFIA21} based on the analyses of \citet{Martins2005, Rauw2005}.
We also accounted for cosmic microwave background ($T=2.7$ K, $U_\textrm{CMB}=0.25$ eV~cm$^{-3}$).

The IC produced by the primary electrons is shown in Fig. \ref{fig:synch} (red dot-dashed line). One can see it cannot be responsible for the observed X-ray non-thermal emission within the model of hadronic gamma-ray emission of Wd2 discussed above -- the IC flux is well below than the detected one. This component does not contribute substantially to the observed gamma-ray emission as well, because the Compton scattering of the optical photons to high-energy gamma-rays is in the Klein-Nishina regime.

However, one cannot exclude the presence in the cluster of some localized lepton populations with energy spectra not extending well beyond 10 MeV, which are able to produce keV IC emission. The low-energy particles suffer strong modulation effect, but they can be accelerated within the winds of massive stars \citep[see e.g.][]{1991ApJ...366..512C} or in the wind collision regions.  
This emission is not expected to be seen in the extended dense arc 
shell structure discussed above. Indeed, the low-energy particles accelerated in the cluster (contrary to VHE ones) have relatively small diffusion coefficient and, as we do know from the solar wind case, suffer from the strong adiabatic losses while propagating through the wind produced by the cluster. The IC component produced by mildly relativistic electron population is likely to be extended above 10-20 keV, which is not expected for the synchrotron component. Further observations of Wd2 in hard X-ray  energy range are required to rule out one of the models.

With some uncertainties, the X-ray data can support the hadronic model of very high energy gamma-ray origin in Wd2. 
The energy requirements for the hadronic emission model are rather modest given the apparent presence of high density material in the shells surrounding Wd2 as observed by \citet{Wd2_shell_SOFIA21}. This is in contrast with the model of gamma-ray emission from 
Westerlund 1 \citep{Wd1_Harer23} where the radiation produced by CRs accelerated at the putative termination shock of the cluster wind  where the plasma density is low. The low density 
environment of the accelerator in this model makes the IC emission more preferable on the energetic ground.

\section{Conclusions}
\label{sec:con}
The compact cluster of young massive stars Westerlund 2 was detected in June-July 2022 with the \art\ telescope in the energy range 4-12 keV and its data were analysed together with the long archival \chandra\ observations in the 0.4-8 keV band. One of the aims of the study was a search for a non-thermal X-ray component from Wd2 since the source was earlier detected in gamma rays and as well as Westerlund 1 it is considered as a perspective source of high-energy cosmic rays \citep{Aha19}.   

The X-ray spectrum of Wd2 and its vicinity extends at least to 12 keV and its fitting requires the presence of two thermal plasma components of electron temperatures about 0.1 and 1 keV and in addition some hard emission. The hard component is either a very hot thin thermal plasma with electron temperature of $\sim$ 5 keV or a non-thermal emission which can be modelled as a power-law component with the photon index of $\sim$ 2. To distinguish between these two alternatives we performed a 3D MHD modeling of the plasma produced by the colliding winds of young massive stars in the cluster simulating Wd2. The model provided plasma temperatures of $\sim~\mathrm{keV}$, but did not indicate a presence of high electron temperatures $>$ 3 keV required for the Wd2 data fitting. This makes the presence of non-thermal component rather likely. The power-law component can be interpreted as a synchrotron radiation of multi-TeV electrons directly accelerated in the cluster core. 

The hard component was also found well outside the cluster core in the projected position of the extended dense arc shell structure driven by stellar cluster winds, which was revealed in the [C II] 158 $\mu$m line supplemented with $^{12}$CO and $^{13}$CO line observations by \citet{Wd2_shell_SOFIA21}. The presence of the non-thermal component in the dense shell can be explained by the synchrotron radiation of either primary electrons from the cluster core or secondary $e^{\pm}$ pairs produced by inelastic hadronic collisions of VHE protons accelerated in the core with the dense matter in the shell. The latter model assumes that the dense gas in the shell is accompanied with highly enhanced magnetic field of magnitudes above hundreds of $\mu$G. 

Alternatively, the IC scattering of intense cluster radiation field by low-energy primary electrons in the vicinity of Wd2 is unlikely to produce the observed X-ray emission, unless there is an additional MeV regime electron population giving the X-ray flux. 
Within the frame of high-energy particle acceleration models of Wd2 \citep[see e.g.][]{Wd2_Mestre21,BK22} one can expect that the synchrotron photons cannot have energies well above of a few tens of keV, while the spectrum produced by the IC scattering could be extended up to gamma rays. Sensitive hard X-ray - soft gamma-ray  observations can help to resolve the issue.

\section*{Acknowledgements}
This work is based on observations with  Mikhail Pavlinsky \art\ telescope, hard X-ray instrument on board the \srg\  observatory. The \srg\ observatory was created by Roskosmos (the Lavochkin Association and its subcontractors) in the interests of the Russian Academy of Sciences represented by its Space Research Institute (IKI) in the framework of the Russian Federal Space Program, with the participation of Germany. 
The authors thank Prof. Ruizhi Yang for helpful comments. 
We acknowledge the use of of data provided by NASA ADS system and SIMBAD database, operated at CDS, Strasbourg, France. This research made use of PLUTO public MHD code developed by A. Mignone and the PLUTO team. The system modeling was performed by D.V.B., M.E.K. and A.M.B. at the JSCC RAS and the 'Tornado' subsystem of the St.~Petersburg Polytechnic University super-computing centers with a support from RSF 21-72-20020 grant. The analysis of joint observational data was performed by Yu. A. Uvarov under support of    
the baseline project 0040-2019-0025.

\section*{Data Availability}

The output data may be provided upon a reasonable request. 


\bibliographystyle{mnras}
\bibliography{mnras_template} 


\bsp	
\label{lastpage}

\end{document}